\documentclass[prl,twocolumn,showpacs,preprintnumbers,amsmath,amssymb]{revtex4}
\usepackage{graphicx}
\usepackage{dcolumn}
\usepackage{bm}
\usepackage{epsfig}
\usepackage{dcolumn}
\def\be{\begin{equation}}
\def\ee{\end{equation}}
\def\bea{\begin{eqnarray}}
\def\eea{\end{eqnarray}}\def\nn{\nonumber}
\def\gsim{\ \rlap{\raise 2pt\hbox{$>$}}{\lower 2pt \hbox{$\sim$}}\ }
\def\lsim{\ \rlap{\raise 2pt\hbox{$<$}}{\lower 2pt \hbox{$\sim$}}\ }
\def\dslash{\kern-4pt \not{\hbox{\kern-2pt $\partial$}}}
\def\pslash{\not{\hbox{\kern-2pt p}}}

\def\pmue{{{P_{\mu e}} }}
\def\pmumu{{{P_{\mu \mu}} }}
\def\pemu{{{P_{e \mu}} }}

\newcommand{\dcp}{\delta_{CP}}
\newcommand{\nova}{NO$\nu$A}

\begin{document}
\DeclareGraphicsExtensions{.eps,.ps}


\title{Can atmospheric neutrino experiments provide
the first hint of leptonic CP violation?}



\author{Monojit Ghosh}
\affiliation{
Physical Research Laboratory, Navrangpura,
Ahmedabad 380 009, India}
 
\author{Pomita Ghoshal}
\affiliation{
Physical Research Laboratory, Navrangpura,
Ahmedabad 380 009, India}
 
\author{Srubabati Goswami}
\affiliation{
Physical Research Laboratory, Navrangpura,
Ahmedabad 380 009, India}

\author{Sushant K. Raut}
\affiliation{
Physical Research Laboratory, Navrangpura,
Ahmedabad 380 009, India}

\begin{abstract}

The measurement of a non-zero value of
the 1-3 mixing angle has paved the way 
for the determination
of leptonic CP violation.
However the current generation long-baseline experiments 
T2K and \nova\ have limited sensitivity to $\dcp$.
In this paper we show, for the first time, the significance
of atmospheric
neutrino experiments in providing the first hint of 
CP violation in conjunction with T2K and \nova.  
In particular, we find that adding atmospheric neutrino data from 
the ICAL detector at the 
India-based Neutrino Observatory (INO) to 
T2K and \nova results in a two-fold increase in  
the range of $\dcp$ values
for which a $2\sigma$  hint of CP violation  can be obtained.
In fact in the  parameter region  
unfavorable for the latter experiments, 
the first signature of CP violation may  well 
come from the inclusion of  atmospheric neutrino data. 

\end{abstract}
\pacs{14.60.Pq,14.60.Lm,13.15.+g}
\maketitle


\underline{\it{\bf Introduction}}:
CP symmetry refers to invariance under simultaneous transformation
of charge conjugation and parity.
A small violation of this symmetry is observed in the quark sector,
in the decays of K and B mesons \cite{quarkcpv}.
This can be explained by the
complex phases in the CKM matrix.
Thus it is natural to expect that CP violation (CPV) occurs in the
lepton sector as well \cite{leptoncpv}. This is reinforced by the observation of neutrino
oscillations which establishes non-zero masses and mixing of these particles.
The MNSP matrix in such a situation would 
contain complex phases. In the basis where the charged 
leptons do not mix amongst each other this matrix is characterized by
three mixing angles ($\theta_{12}$, $\theta_{13}$, 
$\theta_{23}$) and three phases. 
Oscillation experiments are sensitive to the so-called Dirac phase 
$\delta_{CP}$.
This could also be linked to the 
origin of  the observed 
matter-antimatter asymmetry of the universe through
the mechanism of leptogenesis \cite{leptogenesis}. 
Thus, understanding the origin of CPV is one of the 
central themes in particle physics and cosmology. 

Global fits of world neutrino data
give the best-fit values for the neutrino
oscillation parameters (and their $1\sigma$ ranges) as 
$\sin^2 \theta_{12} = 0.31 \pm 0.02$,
$\sin^2 \theta_{23} = 0.39 \pm 0.02$, 
$|\Delta_{31}| = (2.43 \pm 0.1) \times 10^{-3}$ eV$^2$,
$\Delta_{21} = (7.54 \pm 0.26) \times 10^{-5}$ eV$^2$
~\cite{global}.
Here $\Delta_{ij} = m_i^2 - m_j^2$ denotes
the  mass-squared differences.
Recently the third mixing angle $\theta_{13}$ 
has been measured by 
reactor ~\cite{reactors} 
and accelerator ~\cite{eps2013t2k}
neutrino experiments and
the best-fit 
value is  
$\sin^2 2\theta_{13} \approx 0.10 \pm 0.01$~
\cite{global}.
Thus, the remaining unknown quantities are
the neutrino mass hierarchy
(normal hierarchy (NH): $\Delta_{31} > 0$ or inverted hierarchy (IH): 
$\Delta_{31} < 0$), the octant of
$\theta_{23}$ and the CP-phase $\dcp$.

In the MNSP matrix $\dcp$ is associated with 
$\theta_{13}$. Thus a non-zero  $\theta_{13}$ is
required for any measurement of $\dcp$. The
$10\sigma$ signature for non-zero $\theta_{13}$ 
leads naturally to the question of  whether and to what 
extent CPV discovery is possible by 
the current superbeam experiments T2K and \nova.  
In these experiments, the sensitivity to $\dcp$ comes mainly from the 
$\nu_\mu - \nu_e$ (and $\overline{\nu}_\mu - \overline{\nu}_e$)
oscillation probability, $\pmue$ ($P_{\overline{\mu}\overline{e}}$).
Since the probabilities increase 
with $\theta_{13}$,
the relatively large value of $\theta_{13}$  
is expected to be conducive to the measurement of $\dcp$ 
because of increased statistics.  However,   
the statistical error coming from the $\dcp$-independent 
dominant term in $\pmue$ also increases, which tends to reduce the sensitivity 
\cite{large_t13_bad}.  

In addition, the correct signal can  also be faked by a 
wrong hierarchy-wrong $\dcp$ combination~\cite{degen}. 
This hierarchy-$\dcp$ degeneracy makes the measurement of $\dcp$ 
difficult in practice, and higher intensity and/or longer baseline
beam-based experiments have been proposed \cite{proposals}.  
However till the new experiments dedicated for $\dcp$ are built
one can ask whether the LBL experiments T2K and \nova\
can provide any hint for CP violation. 
 

In \cite{novat2k}, it was shown that a prior knowledge of the hierarchy
facilitates the measurement of $\dcp$ by 
\nova\ and T2K. However, the determination of the hierarchy by \nova\ and T2K itself 
suffers from being dependent on the `true' value of $\dcp$ in nature. For the favorable combinations 
(\{$\dcp \in [-180^\circ,0^\circ]$, NH\} or \{$\dcp \in [0^\circ,180^\circ]$, IH\}), 
\nova\ and T2K will be able to determine the 
hierarchy at 90\% C.L. with their planned runs. 
{ But} their hierarchy determination ability 
and hence their CP sensitivity will be poor if nature has chosen the
unfavorable combinations~\cite{reopt}. 
On the other hand, the hierarchy sensitivity of atmospheric neutrino 
experiments is independent of $\dcp$~\cite{prd2007}. 
Hence, a combination of long-baseline (LBL) 
and atmospheric data can determine the hierarchy for all 
$\dcp$~\cite{atmoshier,gct} values  including the 
adverse ones. 
This can substantially improve the ability of the LBL 
experiments to detect 
CPV in  the unfavorable regions of 
$\dcp$. 
In this paper we  demonstrate that the CP 
sensitivity of T2K and \nova\ can be enhanced significantly 
by including atmospheric neutrino data in the analysis. 
For the latter we consider a magnetized iron calorimeter detector (ICAL) 
which is being developed by the INO collaboration 
\cite{inowebsite}.

 
Since atmospheric neutrino experiments are not sensitive to 
CPV by themselves, their usefulness in determining this property has not been
emphasized so far. 
We show that
for unfavorable values of $\dcp$, atmospheric neutrino 
data from ICAL ameliorates the CPV discovery potential of \nova\ and T2K.
This leads to the possibility of obtaining a $\gsim~ 2\sigma$ hint of CPV 
using
{\it existing and upcoming facilities} 
for a large fraction ($>50\%$) of $\dcp$ values.

\underline{\it{\bf CP violation in neutrino oscillations:}}  
In matter of constant 
density, $\pmue$ can be expressed in terms
of the small parameters $\alpha = \Delta_{21}/\Delta_{31}$ and $s_{13}$
as \cite{akhmedov}
\bea
P_{\mu e }&=
&4 s_{13}^2 s_{23}^2 \frac{\sin^2{[(1 -\hat A)\Delta]}}{(1-\hat A)^2}
\nn \\
&&
+ \alpha \sin{2\theta_{13}}  \sin{2\theta_{12}} \sin{2\theta_{23}} \cos{(\Delta - \dcp)}
\times \nn \\
&&
\frac{\sin{\hat A \Delta}}{\hat A} \frac{\sin{[(1-\hat A)\Delta]}}{(1-\hat A)}
+ {\cal{O}}(\alpha^2) ~,
\label{P-emu}
\eea
where $\Delta = \Delta_{31}L/4E$, 
$s_{ij} (c_{ij}) \equiv \sin{\theta_{ij}}(\cos{\theta_{ij}})$,
$\hat{A} = 2\sqrt{2} G_F n_e E / \Delta_{31}$, 
$G_F$ is the Fermi constant and $n_e$ is the electron number density.
For neutrinos, the signs of $\hat{A}$ and $\Delta$ are positive for NH 
and
negative for IH  and vice-versa for antineutrinos. 
The second term in Eq.~(\ref{P-emu}) is the source of the 
hierarchy-$\dcp$ degeneracy \cite{degen}. 

\underline{\it{\bf Experimental details}}: 
For our study we consider the current generation LBL 
experiments \nova\ and T2K and simulate them using 
the GLoBES package \cite{globes,reopt}.
For \nova, we have assumed a $14$ kT totally active scintillator detector 
and a $0.7$ MW beam running for $5(\nu) + 5(\overline{\nu})$ years. 
Since it is expected to start in 2014, this indicates
a timeline up to about 2024. We have used a re-optimized \nova\ set-up with 
refined event selection criteria \cite{reopt,ryannova}. T2K is 
assumed to have a $22.5$ kT Water \v{C}erenkov detector, and a $0.77$ MW beam 
running for $5(\nu) + 0(\overline{\nu})$ years. Taking into account the current 
lower-power run and proposed upgrades,
this will correspond to a timeline till about 2016. We have checked that a 
T2K run in the neutrino mode alone and a combined neutrino-antineutrino run 
give similar results when combined with \nova. 
For these LBL experiments, we have used the systematic errors and background 
rejection efficiencies used in Ref.~\cite{reopt,ryannova}.

For atmospheric neutrinos we consider 
ICAL@INO, which is 
capable of detecting muon events with charge
identification, with a proposed mass of 50 kT \cite{inowebsite}. 
For our analysis 
we use  neutrino energy and angular resolutions of (10\%,$10^\circ$)   
unless noted otherwise. 
These are representative values which give similar sensitivity as obtained 
in \cite{gct} using energy and angular resolutions of  
muons from INO simulation code.  The details of our atmospheric analysis
can be found in \cite{prd2007}. The detector is expected to be operational by 2018/19. 
We present results for exposures of $500$ ($250$) kT yr,
corresponding to a 10(5)-year run i.e. a timeline till about 2028 (2023-24).
We note that the latter is the expected time frame for
\nova\ to complete a 10-year run.

\underline{\it{\bf CP sensitivity in atmospheric neutrinos}}:
\begin{figure*}[t]
\begin{tabular}{rl}
\epsfig{file=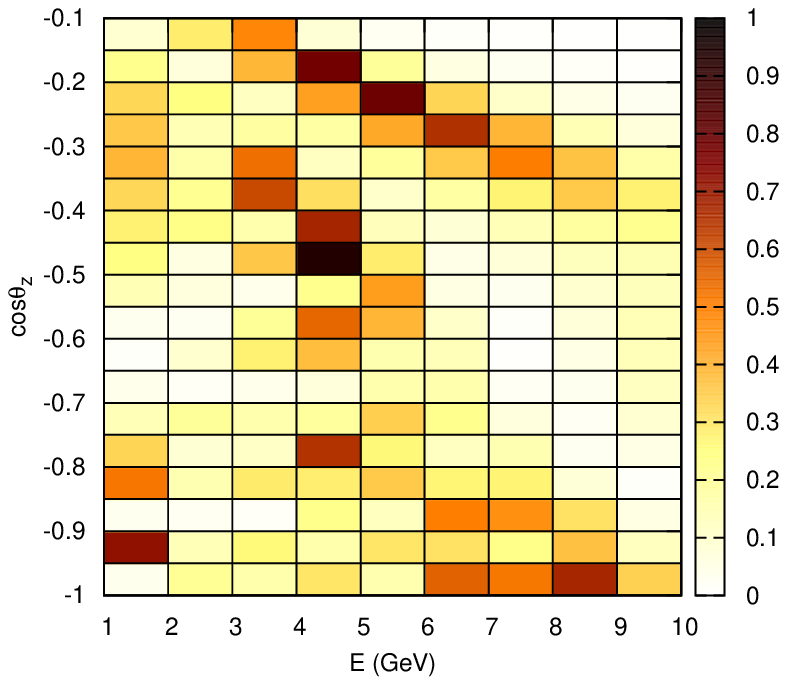,width=8cm,height=6cm,bbllx=100, bblly=60, bburx=350, bbury=280,clip=} &
\epsfig{file=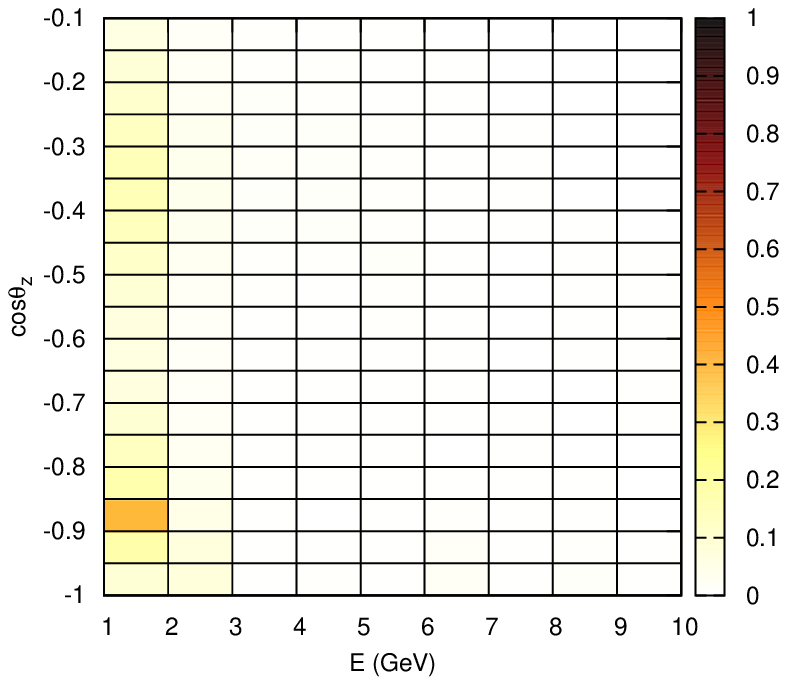,width=8cm,height=6cm,bbllx=100, bblly=60, bburx=350, bbury=280,clip=} 
\end{tabular}
\vspace*{-4ex}
\caption{$S_\mu + S_{\overline{\mu}}$, a measure of ICAL $\dcp$-sensitivity in the 
$E-\cos {\theta_z}$ plane for 
$\sin^2 2\theta_{13} = 0.1$, $\sin^2 \theta_{23} = 0.5$ and NH. 
The grid represents bins in energy and $\cos\theta_z$.
The left panel is with ideal detector resolution and the right panel is with a resolution of $10^\circ$ in angle and $10\%$ in energy.
}
\label{nodcp}
\end{figure*}
The muon events in atmospheric neutrinos get contributions from both  
$\pmumu$ and $\pemu$.
In these probabilities, the $\dcp$-dependent 
term always appears along with a factor of $\cos\Delta$ or $\sin\Delta$. 
If we consider even a $10\%$ error range in
the zenith angle and 
energy of the neutrino, this oscillating term varies over an entire cycle 
in this range. 
As a result, the $\dcp$-sensitivity of the channel gets washed out because of smearing. 
In Fig.~\ref{nodcp}, we have plotted the quantity $S = S_\mu + S_{\overline{\mu}}$ in 
the $\cos\theta_{z}-E$ plane, which 
is a measure of the $\dcp$-sensitivity of the atmospheric neutrino experiment. Here, 
$S_\mu = (\delta N_\mu)^2 / N_\mu(\textrm{avg})$ ~,
where $\delta N_\mu$ is the maximum difference in events
obtained by varying $\dcp$, 
and $N_\mu(\textrm{avg})$ is the average number of events over all values of $\dcp$ 
(and likewise $S_{\overline{\mu}}$ for $\overline{\mu}$ events). The quantity $S$ is 
thus a measure of the 
maximum possible relative variation in events due to $\dcp$ in each bin. 
In the left panel, we show the results 
for an ideal detector with an exposure of 500 kT yr, with infinite energy and angular 
precision. 
Here we see substantial sensitivity to $\dcp$, with $S$ exceeding $0.5$ 
in some bins \cite{samanta}. 
However, when we introduce realistic resolutions 
($10^\circ$ in angle and $10\%$ in energy), we  
see in the right panel that the sensitivity is lost.
This is mostly due to the effect of angular smearing. 
Thus atmospheric neutrino experiments by themselves are not sensitive to 
$\dcp$.
For beam experiments, since the direction of the neutrinos 
is known, angular smearing is not needed and hence 
the sensitivity to $\dcp$ is not
compromised due to this reason.

\begin{figure*}[t]
\includegraphics[width =8cm,height=5.3cm]{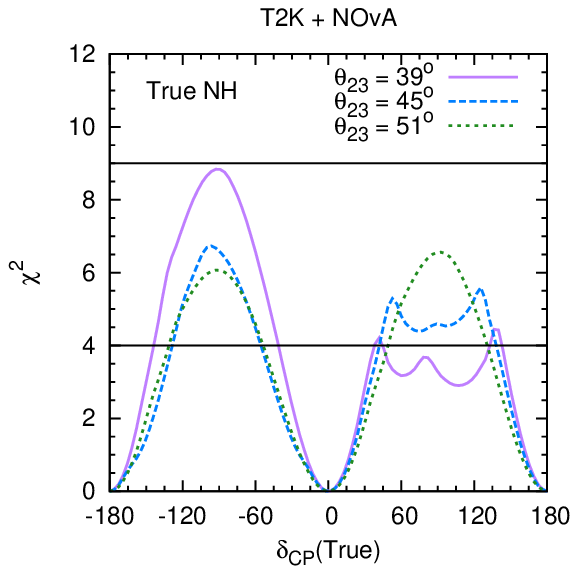}
\includegraphics[width =8cm,height=5.3cm]{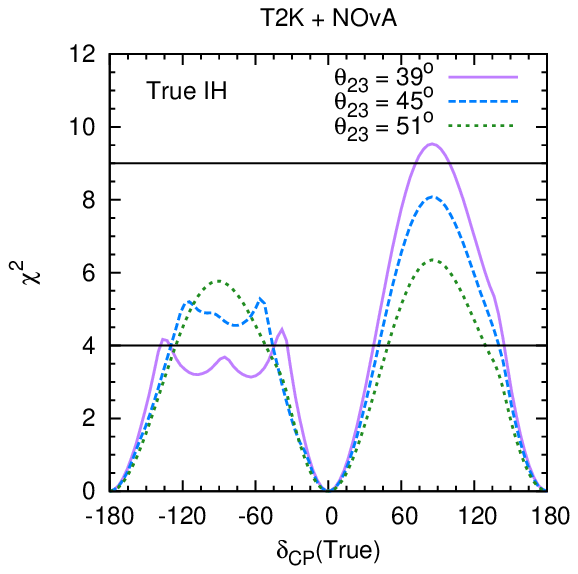} \\
\includegraphics[width =8cm,height=5.3cm]{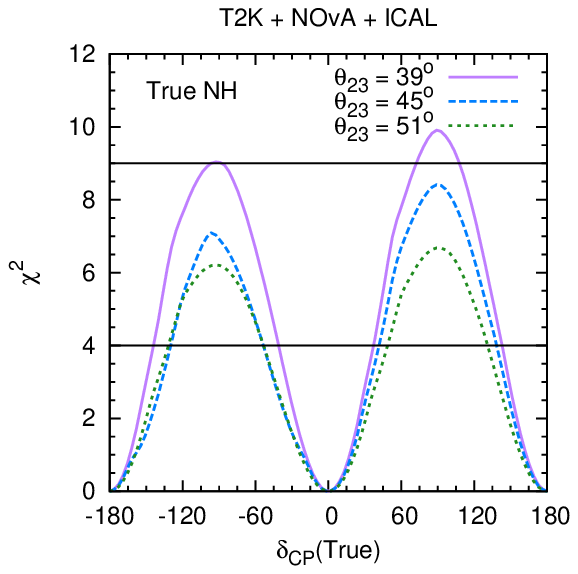}
\includegraphics[width =8cm,height=5.3cm]{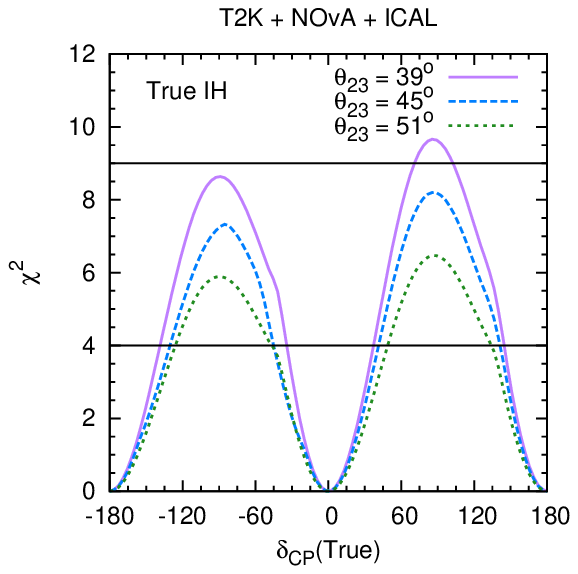}
\vspace*{-4ex}
\caption{CPV discovery vs true $\dcp$ for \nova+T2K
(upper row) and \nova+T2K+ICAL (lower row), for three values
of $\theta_{23}$, $\sin^2 2\theta_{13} = 0.1$ and a true normal (left panels) or 
inverted (right panels) mass hierarchy, {{with 500 kT yr exposure for ICAL}}.
}
\label{discovery}
\end{figure*}

\underline{\it{\bf CP violation discovery}}:
The discovery potential 
for CPV is computed by 
considering a variation of $\dcp$ over the full range $[-180^\circ,180^\circ)$ in the 
simulated true event spectrum,
and comparing this with $\dcp = 0^\circ$ or $180^\circ$ in the test event spectrum. 
The statistical $\chi^2$ for our analysis is defined as 
\begin{equation}
 \chi^2_{stat} = \sum_{bins} \frac{(N^{true}(\dcp) - N^{test}(\dcp=0,\pi))^2}{N^{true}(\dcp)} ~.
\end{equation}
We have accounted for systematic errors by using the method of pulls. For a particular 
value of $\dcp$ in the true spectrum, the resulting $\chi^2_{stat+syst}$ is evaluated for 
test $\dcp = 0$ and $\pi$ and test hierarchy NH and IH. We also marginalize 
over the atmospheric parameters and $\sin^2 2\theta_{13}$. The minimum over all these 
test parameter combinations is chosen as the final $\chi^2$.
This is then studied as a function of (true) $\dcp$. 
In Fig.~\ref{discovery}, we plot the CPV discovery potential of the 
LBL experiments
\nova\ and T2K and the ICAL detector with 500 kT yr exposure.
The upper panels give the CP discovery for the combination \nova+T2K, while 
the lower panels depict the results
for \nova+T2K+ICAL. The true neutrino hierarchy is assumed to be 
NH(IH) in the left(right) column. 

From the figure, it may be observed that the CPV discovery of \nova+T2K 
suffers a drop in one of the half-planes 
of $\dcp$, depending on the true hierarchy - in the region 
$[0^\circ,180^\circ]$ if it is NH,
and $[-180^\circ,0^\circ]$ if it is IH. This is due to the fact 
that the hierarchy sensitivity
of \nova+T2K is highly sensitive to $\dcp$, and becomes low in the 
unfavorable regions \cite{novat2k}.
Consequently, for unfavorable $\dcp$ values, 
marginalization over the hierarchy causes 
the \nova+T2K CPV discovery potential to drop, 
since the minimum for CPV discovery can then occur in 
conjunction with the wrong hierarchy.  

However, an atmospheric neutrino detector like ICAL gives a 
hierarchy sensitivity 
which is remarkably
stable over the entire range of $\dcp$, 
even though it does not offer any significant 
CPV discovery potential by itself.
This hierarchy sensitivity 
excludes the wrong-hierarchy minimum for CPV discovery. 
When this information is added to \nova+T2K, the drop in the CPV discovery 
in the unfavorable half-planes of $\dcp$ is resolved.

The results depend significantly on the true value of 
$\theta_{23}$.  
In the favourable   
$\dcp$ region the discovery potential becomes worse with increasing $\theta_{23}$.
This is 
because the  $\dcp$-independent leading term in Eq. \ref{P-emu} increases with $\theta_{23}$,  
giving a higher statistical error,   
while the CP dependent term has
only a weak dependence on this parameter \cite{hubercpv}. 
In the unfavorable region since the $\chi^2$ minimum comes with the wrong hierarchy, 
the dependence of hierarchy sensitivity on $\theta_{23}$ also comes into play.
This   
causes a drop in the value of $\chi^2$ for lower values of $\theta_{23}$. 

The advantage offered by combining ICAL with the LBL data is most prominent for
$\theta_{23} = 39^\circ$,
and progressively diminishes with increasing $\theta_{23}$. 
For $\theta_{23} = 39^\circ$ or $45^\circ$
the
hierarchy sensitivity of \nova+T2K is poorer, with the minimum occuring
for the
wrong hierarchy. This can be ruled out by atmospheric neutrino data.
For $\theta_{23} = 51^\circ$ the ICAL information is
superfluous,  
since 
the hierarchy sensitivity
of the \nova+T2K combination itself is good enough
to exclude the wrong hierarchy CPV discovery minima
even for unfavorable $\dcp$ values. 
In general, the atmospheric neutrino contribution to the CPV discovery
potential of \nova+T2K+ICAL
is effective till the wrong hierarchy solution is disfavored and the minimum 
comes with the true hierarchy. Once that is achieved, a further increase in hierarchy 
sensitivity of atmospheric neutrinos will not affect the CPV discovery results, 
since atmospheric neutrinos by themselves do not have CPV sensitivity 
for realistic resolutions. 

To quantitatively understand the significance of the atmospheric neutrino contribution,
we consider true NH and $\theta_{23} = 39^\circ$ ($45^\circ$). In this case,
the ICAL contribution required to exclude the wrong hierarchy CPV minima
is about $\chi^2 = 6.5 (4)$. Thus a hierarchy sensitivity of
$\sim 2.5 (2) \sigma$
is enough to rule out
the wrong hierarchy solutions and compensate for the drop in CPV discovery
potential of \nova+T2K in the
unfavorable $\dcp$ region.
For true NH(IH), T2K+\nova\ can  discover CPV at $2\sigma$ for 
$\sim$ 32\%(35\%) fraction of $\dcp$ values for $\theta_{23}= 39^\circ$. 
By adding ICAL information, this improves to $\sim$ 58\%. 
For maximal CPV ($\dcp = \pm 90^\circ$), inclusion of 
ICAL gives a $\sim 3\sigma$ signal for both hierarchies. 
Without the ICAL { contribution} this is true only in one of the half-planes 
depending on the hierarchy. 

\begin{figure}
\includegraphics[width =8cm,height=5.3cm]{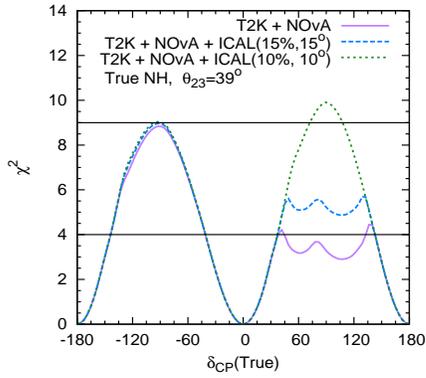}
\vspace*{-4ex}
\caption{CPV discovery vs true $\dcp$ for \nova+T2K and 
\nova+T2K+ICAL {{(500 kT yr)}} for two sets of ICAL detector resolutions for $\theta_{23}=39^\circ$, 
$\sin^2 2\theta_{13} = 0.1$ and true NH.}
\label{resvary}
\end{figure}

{To study the effect of ICAL detector resolutions on the results}, we plot in 
Fig.~\ref{resvary} the CPV discovery potential of \nova+T2K+ICAL for 
$\theta_{23} = 39^\circ$ and true NH assuming two sets of energy and angular 
smearings for ICAL - 
(15\%,$15^\circ$) and (10\%,$10^\circ$).
In the former case, although an indication of  CPV at  
$2\sigma$ is seen to be 
achieved in the unfavorable half-plane,
the $\chi^2$ minima still occur with the wrong hierarchy, as evident from the 
shape of the curve. For the wrong hierarchy the shape is 
dictated by a mismatch in both $\delta_{CP}$ and hierarchy between the true and test 
event spectra, and one does not get a smooth dependence on $\delta_{CP}$. 
For the latter (better) smearing set, the $\chi^2$ minimum shifts to the 
true-hierarchy regime and the sensitivity comes only from $\delta_{CP}$. Hence the smooth 
behavior of the curve is restored. An improvement in the resolution beyond 
(10\%,$10^\circ$)
would not affect the CPV discovery potential significantly (with the same exposure), since the minima already occur 
with the true hierarchy where the atmospheric neutrino contribution is negligible. 
However, for a superior angular resolution there can be a slight increase 
in the CPV discovery $\chi^2$ coming from atmospheric neutrinos themselves.

\begin{figure}
\includegraphics[width =8cm,height=5.3cm]{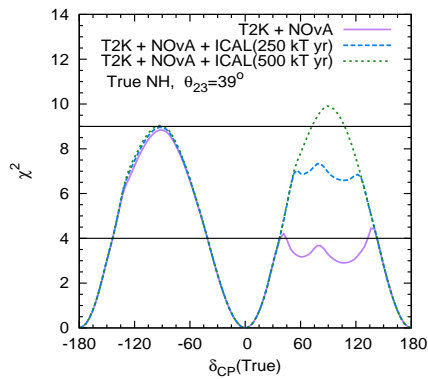}
\vspace*{-4ex}
\caption{CPV discovery vs true $\dcp$ for \nova+T2K and 
\nova+T2K+ICAL for two exposures, 250 kT yr and 500 kT yr for $\theta_{23}=39^\circ$, 
$\sin^2 2\theta_{13} = 0.1$ and true NH. The ICAL resolutions are assumed to be 
10$\%$ in energy and 10$^\circ$ in angle.}
\label{exposure}
\end{figure}

In order to gauge the contribution from ICAL with a reduced exposure, we plot 
in Fig.~\ref{exposure} the CPV discovery as a function of $\dcp$ for \nova+T2K and 
\nova+T2K+ICAL for two ICAL exposures, 250 kT yr and 500 kT yr for $\theta_{23}=39^\circ$, 
$\sin^2 2\theta_{13} = 0.1$ and true NH, using the (10\%,$10^\circ$) ICAL resolution set.
The figure shows that even though the hierarchy-$\dcp$ degeneracy is not fully resolved 
with an ICAL exposure of 250 kT yr, 
a 2.5$\sigma$ hint for CPV is still 
achieved over a large part of the unfavorable half-plane even with this 
exposure, 
corresponding to a 5-year run till 
2023/2024. Hence a chronologically matched run-time of \nova\ and ICAL is still conducive 
to a significant gain in giving a hint of  CPV 
when ICAL is combined.
However, irrespective of the time scale of the 
different experiments, for  
parameter values in the unfavorable half-plane, the first 
signature of CPV may come after adding atmospheric neutrino data
to T2K/\nova.

\underline{\it{\bf Conclusions}}: 
In this paper we emphasize the critical impact that 
atmospheric neutrinos can have in obtaining the first hint of  CPV   
from the LBL experiments T2K/\nova.  This is achieved by the  
ability of the atmospheric neutrino data to exclude the 
degenerate wrong-hierarchy solutions. Taking ICAL@INO as the 
representative detector, we show that adding this data 
to T2K and \nova\ can provide a signature of  
CPV at $2\sigma$ for almost twice the 
range of $\dcp$ values ($\sim 58\%$). For maximal CPV 
the significance of the signal can reach $3\sigma$ in the unfavorable 
half-plane also. The effect of adding ICAL data is more prominent if 
$\theta_{23}$ is in the lower octant, where the current best-fit value 
lies.  
In fact, if nature has chosen such unfavorable
combinations of parameters then it is the addition of 
atmospheric neutrino data to T2K+\nova\ which may give us the 
first signal of CPV.

We note that the idea discussed in this paper can be of importance 
and interest to other atmospheric and/or reactor 
experiments sensitive to the 
mass hierarchy and  can  initiate 
similar studies. 
This aspect should also be taken into account                               
in planning strategies          
for future experiments to measure CPV more precisely
\cite{euroical}. 

\noindent\underline{\it{\bf Acknowledgements}}:
We thank A. Dighe, R. Gandhi and S. Uma Sankar for useful comments and discussions.


\end{document}